\documentclass[preprint]{aastex61}

\shorttitle{Association between tornadoes and hosting prominence instability}
\shortauthors{Mghebrishvili et al.}

\begin{document}

\title{Association between tornadoes and instability of hosting prominences}


\author{Irakli Mghebrishvili}
\affiliation{Abastumani Astrophysical Observatory at Ilia State University, 3/5 Cholokashvili Avenue, 0162, Tbilisi, Georgia, E-mail: irakli.mghebrishvili.1@iliauni.edu.ge}

\author{Teimuraz V. Zaqarashvili}
\affiliation{Space Research Institute, Austrian Academy of Sciences, Schmiedlstrasse 6, 8042 Graz, Austria}
\affiliation{Abastumani Astrophysical Observatory at Ilia State University, 3/5 Cholokashvili Avenue, 0162, Tbilisi, Georgia, E-mail: irakli.mghebrishvili.1@iliauni.edu.ge}
\affiliation{IGAM-Kanzelh\"ohe Observatory, Institute of Physics, University of Graz, Universit\"atsplatz 5, 8010 Graz, Austria}

\author{Vasil Kukhianidze}
\affiliation{Abastumani Astrophysical Observatory at Ilia State University, 3/5 Cholokashvili Avenue, 0162, Tbilisi, Georgia, E-mail: irakli.mghebrishvili.1@iliauni.edu.ge}

\author{David Kuridze}
\affiliation{Institute of Mathematics, Physics and Computer Science, Aberystwyth University, Ceredigion, Cymru, SY23 3, UK}
\affiliation{Abastumani Astrophysical Observatory at Ilia State University, 3/5 Cholokashvili Avenue, 0162, Tbilisi, Georgia, E-mail: irakli.mghebrishvili.1@iliauni.edu.ge}
\affiliation{Astrophysics Research Centre, School of Mathematics and Physics, Queen�s University Belfast, Belfast BT7 1NN, UK}

\author{David Tsiklauri}
\affiliation{School of Physics and Astronomy, Queen Mary University of London, London, E1 4NS, United Kingdom}

\author{Bidzina M. Shergelashvili}
\affiliation{Space Research Institute, Austrian Academy of Sciences, Schmiedlstrasse 6, 8042 Graz, Austria}
\affiliation{Abastumani Astrophysical Observatory at Ilia State University, 3/5 Cholokashvili Avenue, 0162, Tbilisi, Georgia, E-mail: irakli.mghebrishvili.1@iliauni.edu.ge}
\affiliation{Combinatorial Optimization and Decision Support, KU Leuven campus Kortrijk, E. Sabbelaan 53, 8500 Kortrijk, Belgium}

\author{Stefaan Poedts}
\affiliation{Center for Mathematical Plasma Astrophysics, Department of Mathematics, KU Leuven, 200 B, B-3001, Leuven, Belgium}

\begin{abstract}
We studied the dynamics of all prominence tornadoes detected by the \emph{Solar Dynamics Observatory}/Atmospheric Imaging Assembly from 2011 January 01 to December 31. In total, 361 events were identified during the whole year, but only 166 tornadoes were traced until the end of their lifetime. Out of 166 tornadoes, 80 (48\%) triggered CMEs in hosting prominences, 83 (50\%) caused failed coronal mass ejections (CMEs) or strong internal motion in the prominences, and only 3 (2\%) finished their lifetimes without any observed activity. Therefore, almost all prominence tornadoes lead to the destabilization of their hosting prominences and half of them trigger CMEs. Consequently, prominence tornadoes may be used as precursors for CMEs and hence for space weather predictions.

\end{abstract}

\keywords{Sun: atmosphere -- Sun: filaments -- Sun: prominences -- Sun: oscillations}

\section{Introduction} \label{sec:intro}

Tornadoes are rotating, vertical structures frequently observed in the Earth's atmosphere. The term "tornado" has also been used to define observed magnetic structures in the solar atmosphere. There are two types of solar tornadoes. One type is connected with small-scale chromospheric swirls or chromospheric tornadoes, which are observed in chromospheric spectral lines and may provide a mechanism for channeling energy from the lower into the upper solar atmosphere \citep{Wedemeyer2012}. Another type is related with larger structures typically associated with solar filaments/prominences and is referred to as prominence tornadoes. Prominence tornadoes were first observed almost a century ago by \citet{Pettit1925} as a closely twisted rope or a fine screw. The high spatial and temporal resolution full-disk images obtained by the Atmospheric Imaging Assembly (AIA; \citet{Lemen2012}) on board the \emph{Solar Dynamics Observatory} (\emph{SDO;} \citet{Pesnell2012}) revived interests in their structure. Prominence tornadoes are usually seen in hot coronal lines (e.g.  171 {\AA}) as thin dark vertical structures at the solar limb \citep{Wedemeyer2013}.

Chromospheric tornadoes are considered to be rotating structures, but the rotation of prominence tornadoes is currently under debate. Animations in extreme ultraviolet spectral lines and some spectroscopic measurements show rotation evidence of prominence tornadoes \citep{Suarez2012,Wedemeyer2013, Su2014, Levens2015}. However, recent spectroscopic observations did not confirm the rotational dynamics \citep{Schmieder2017}. Instead, it was suggested that the rotation (if it exists) of prominence tornadoes is only intermittent, lasting no more than one hour \citep{Gonzalez2016}. \citet{Panasenco2014} argued that the elusive rotation of a tornado can be created by oscillations along the prominence spine and/or 3D plasma motion following the magnetic fields inside and along the prominence. The oscillation patterns in tornadoes were frequently observed in recent years \citep{Mghebrishvili2015,Gonzalez2016,Schmieder2017}; therefore, the patterns may create the illusion of rotation. If the prominence tornadoes do not rotate, then the term "tornado" is clearly misleading. However, we will still use the term "prominence tornado" in this paper.

Prominence tornadoes usually appear near prominence legs and may play a significant role in the supply of mass and twists \citep{Su2012,Wedemeyer2013}. The structures are very dynamic during their evolution, sometimes splitting into thin threads and rejoining afterward. On the other hand, prominences/filaments often undergo large-scale instabilities that are usually associated with flares and coronal mass ejections (CMEs; \citet{Labrosse2010,Mackay2010}). The very dynamic behavior of the tornado-like structures naturally suggests their connection with the prominence instability. The tornadoes may become unstable due to some sort of magnetic (or flow) instability, which may destabilize the host prominence on large-scales and lead to flare/CME \citep{Wedemeyer2013}. If the tornadoes trigger an instability in the associated prominence, then they can be used as CME precursors and hence could significantly improve the space weather predictions.

In this paper, we analyze the data obtained by the \emph{SDO}/AIA instrument during the whole year of 2011 to study the dynamics of all prominence tornadoes, which appeared during this time interval. We identify over 361 such structures and are able to obtain statistical information regarding their dynamics. In particular, we investigate the evolution and stability of each tornado, their connection with the dynamics of associated prominences and explore their role in triggering CMEs.

\section{Observations and data analysis} \label{sec:data}

The observations were obtained between 2011 January 1, and December 31 with the SDO. Our goal was to identify all prominence tornadoes during this time and to follow their time evolution individually. Prominence tornadoes are not included in the Heliophysics Events Knowledgebase (HEK); therefore, to observe all prominence tornadoes above the solar disk, we were constrained to do it by naked eye. Throughout the analysis, we made use of JHelioviewer \citep{Mueller2009}, visualization software for solar physics data based on the JPEG 2000 image compression standard for effective viewing and exploration of AIA data. It allowed us to observe the tornadoes in our preferred filters, time interval, and cadence. We defined the starting coordinates and lifetimes of each prominence tornado during 2011. We have measured each tornado's lifetime as the time interval between its first appearance and complete disappearance. AIA 171 {\AA} images were used for this purpose as the prominence tornadoes are expected to be visible as dark structures in this line. Because prominences and associated tornadoes are long lived structures, the 2 hr cadence was selected for the 171 {\AA} line. After determining the final stages of evolution for individual tornadoes, we downloaded level 1 data and processed it to corresponding level 1.5 fits files of the 171, 193, and 304 {\AA} band channels. The level 1.5 data include bad-pixel removal, de-spiking, flat-fielding, and scale correction. The data were calibrated and analyzed using standard routines in the SolarSoft (SSW) package. The final stages of tornado evolution were carefully studied for each event.

Solar tornadoes are mostly visible at the solar limb as dark structures in hot coronal lines. When tornadoes move toward the solar disk due to the rotation, their identification is complicated. First, the hosting prominences also appear as dark absorption filaments on the disk; therefore, it is hard to distinguish tornadoes from them. Second, the tornadoes seem to be vertical structures, and hence their detection on the disk is complicated due to the projection effects. We traced all prominence tornadoes appearing at the west limb from the far side of the Sun until they were seen on the disk. We also traced all tornadoes at the east limb until they disappeared to the far side of the Sun due to the solar rotation.

\begin{figure}
\epsscale{0.59}
\plotone{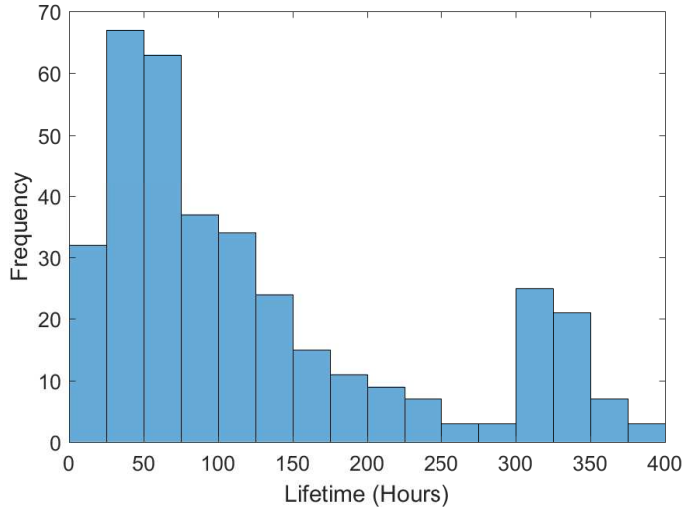}
\caption{Lifetime distribution of all detected prominence tornadoes during the year 2011.\label{fig1}
}
\end{figure}

All information about detected and investigated tornadoes, including starting coordinates, the start and end times, lifetime, and fate (i.e.\ how did tornadoes end their evolution), is presented in a catalog (see Table 2 in the Appendix). During one year, we detected 361 tornadoes. Tornadoes sometimes appeared as separate structures, sometimes they were appearing as a group of different tornadoes. They were also associated with hedgerow prominences \citep{Suarez2012}. In our case, we observed 177 groups of tornadoes, which were composed of two or more tornadoes. Each group was associated with an individual prominence leg; therefore, we considered each group as a single event. Sometimes prominence tornadoes were associated with coronal cavities \citep{li2012, Panesar2013, Wedemeyer2013, Mghebrishvili2015}, which usually appear as dark semicircular or circular regions in the corona above prominences. We identified 113 coronal cavities above prominence tornadoes.

\begin{figure}
\epsscale{0.37}
\plotone{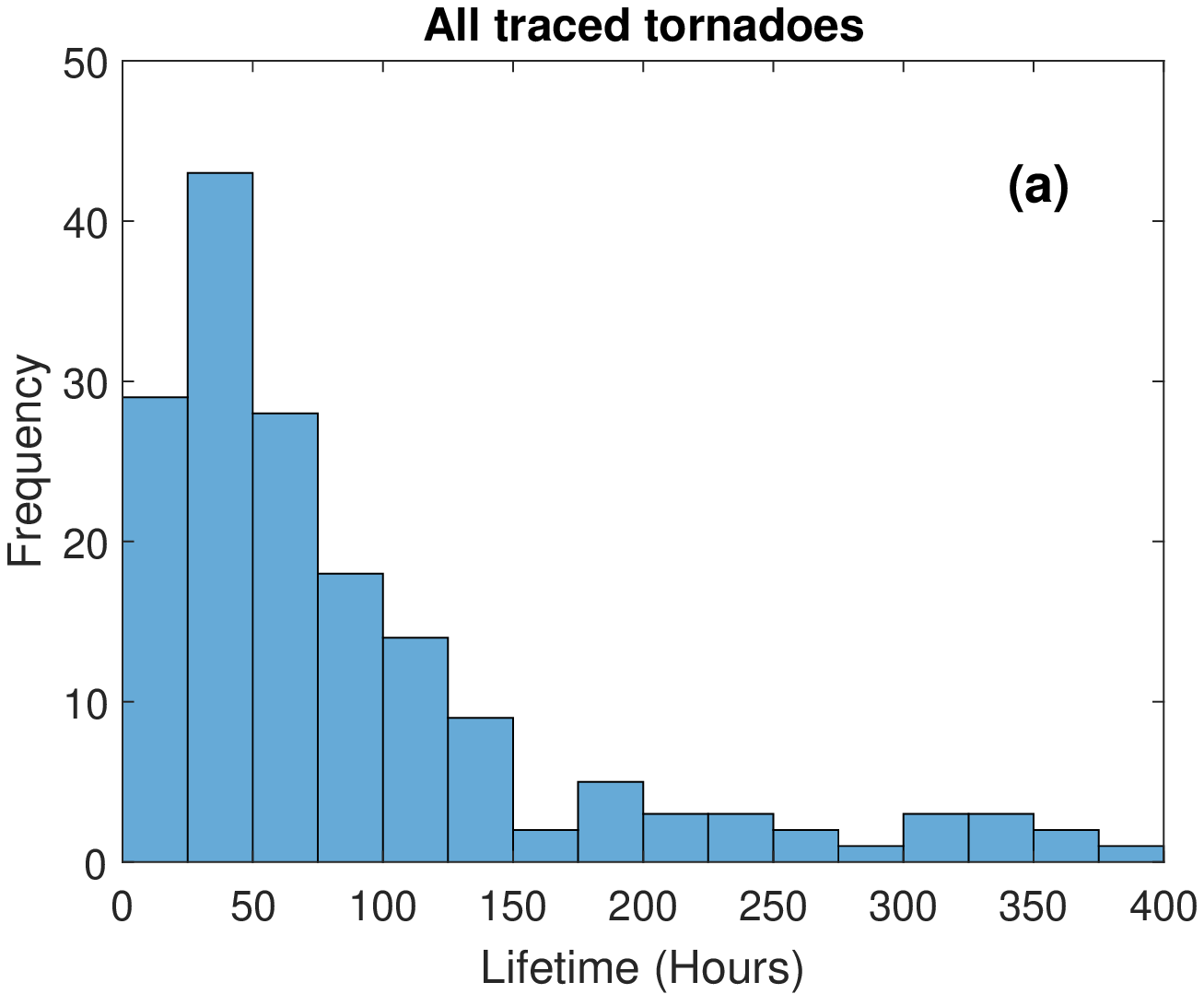}
\plotone{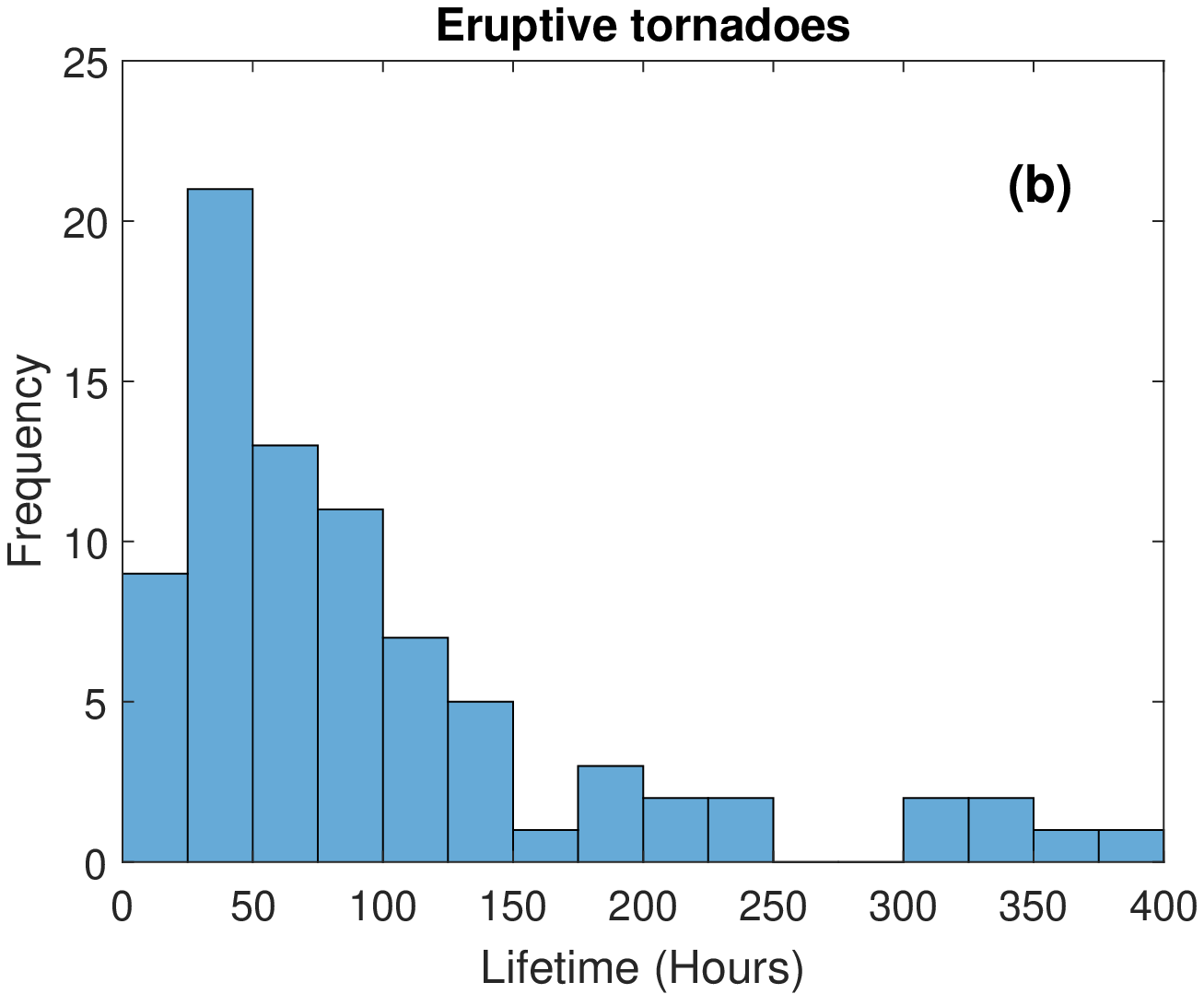}
\plotone{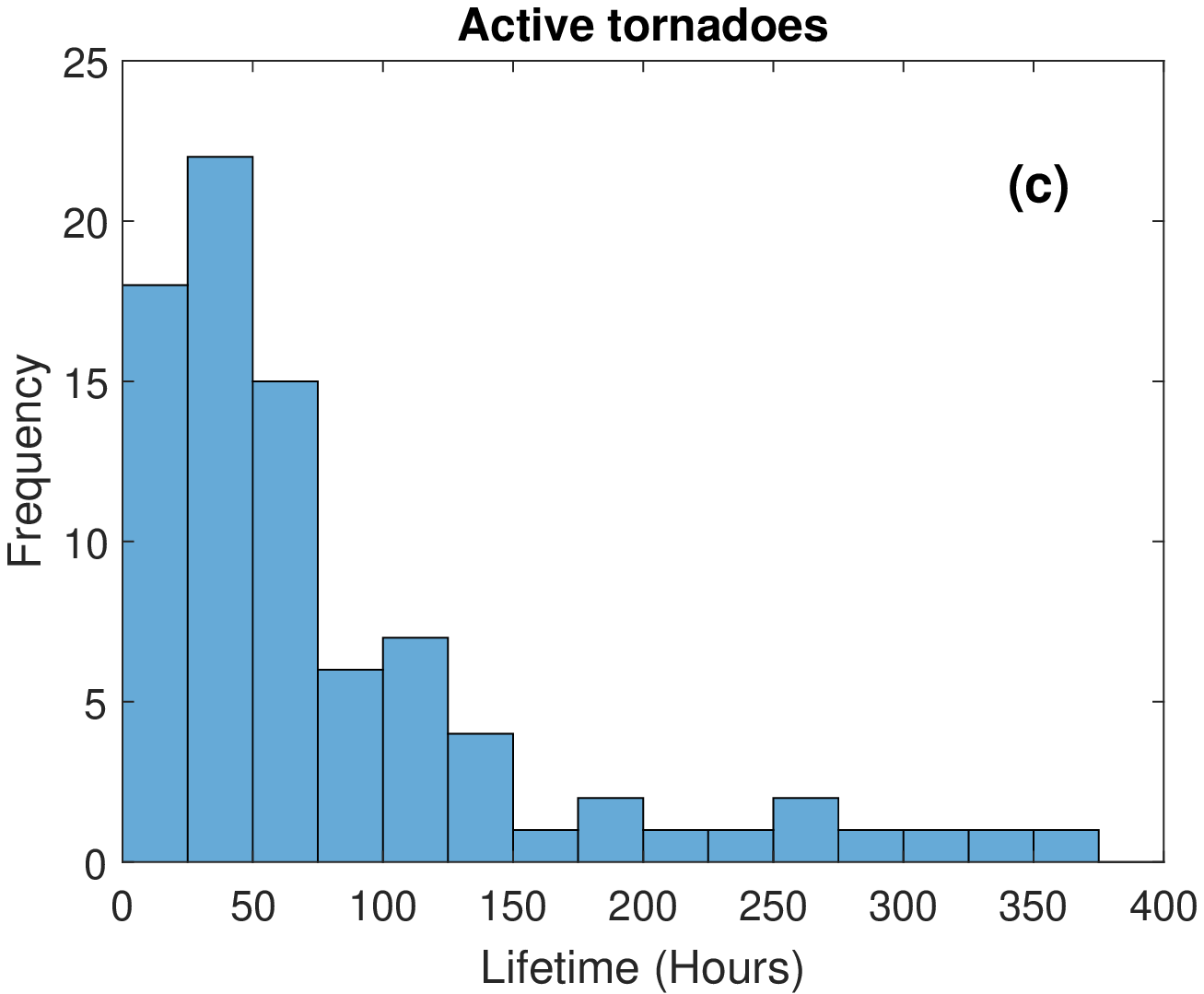}
\caption{Lifetime distribution: (a) all traced prominence tornadoes from the formation to the end of their lifetime, (b) eruptive tornadoes, and (c) active tornadoes.\label{fig2}
}
\end{figure}

\begin{deluxetable}{lc}
\caption{Classification of prominence tornadoes during the year 2011.\label{tab1}}
\tablehead{
\colhead{Category} & \colhead{Number}
}
\startdata
Eruptive tornadoes & 80 \\
Active tornadoes & 83 \\
Quasi-quiescent tornadoes & 3 \\
\enddata
\end{deluxetable}

\begin{figure}
\epsscale{0.57}
\plotone{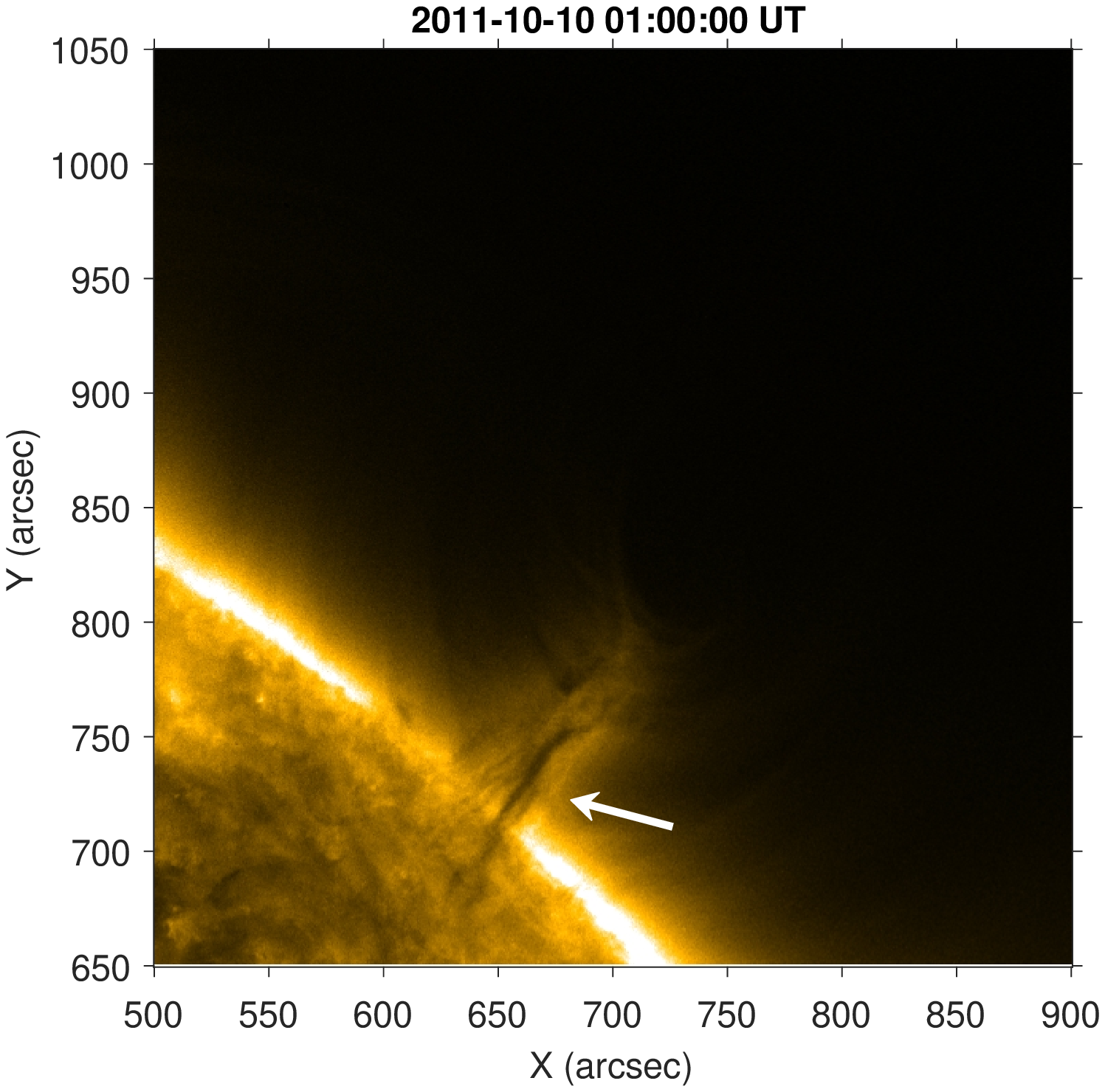}
\plotone{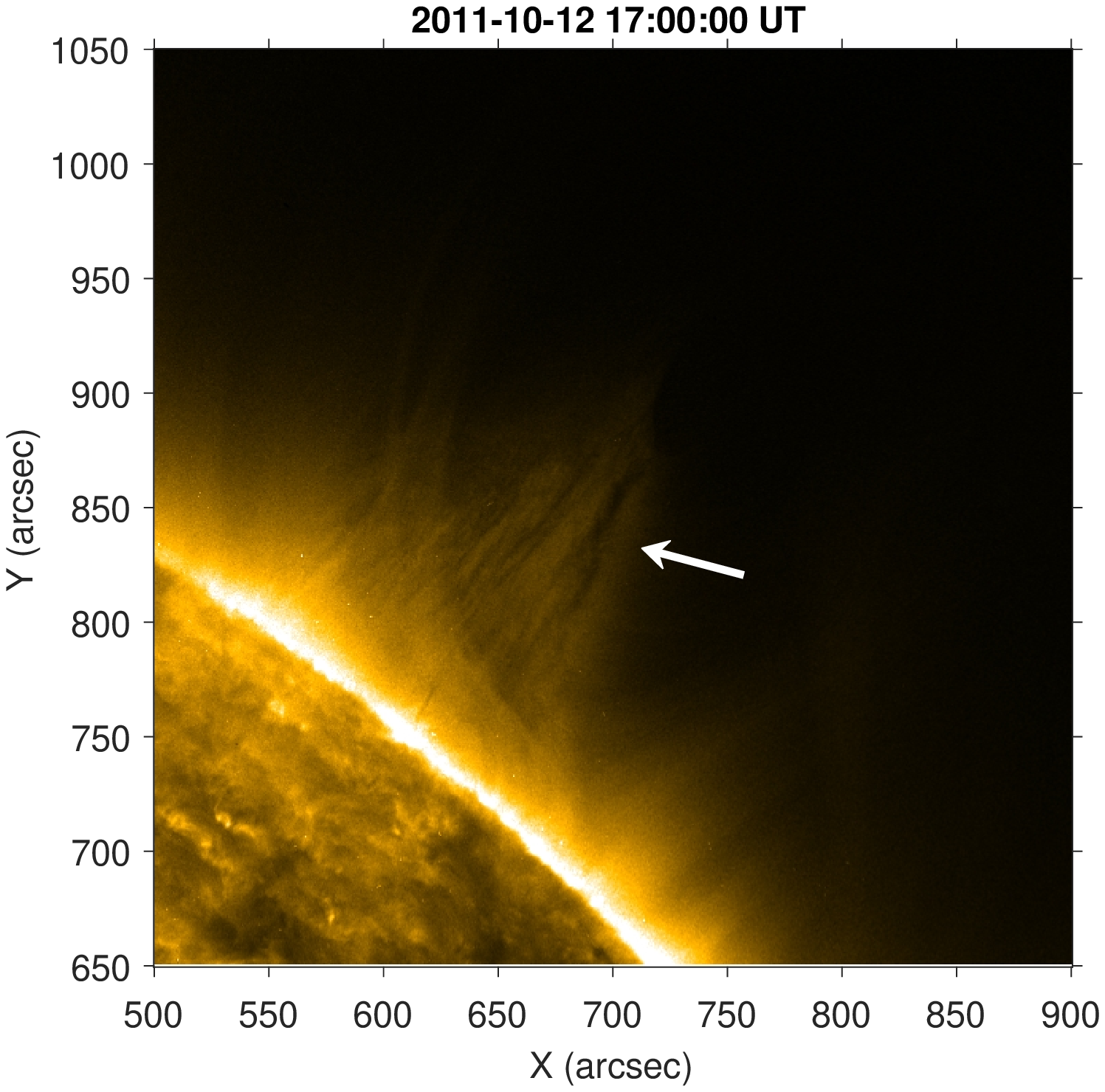}
\plotone{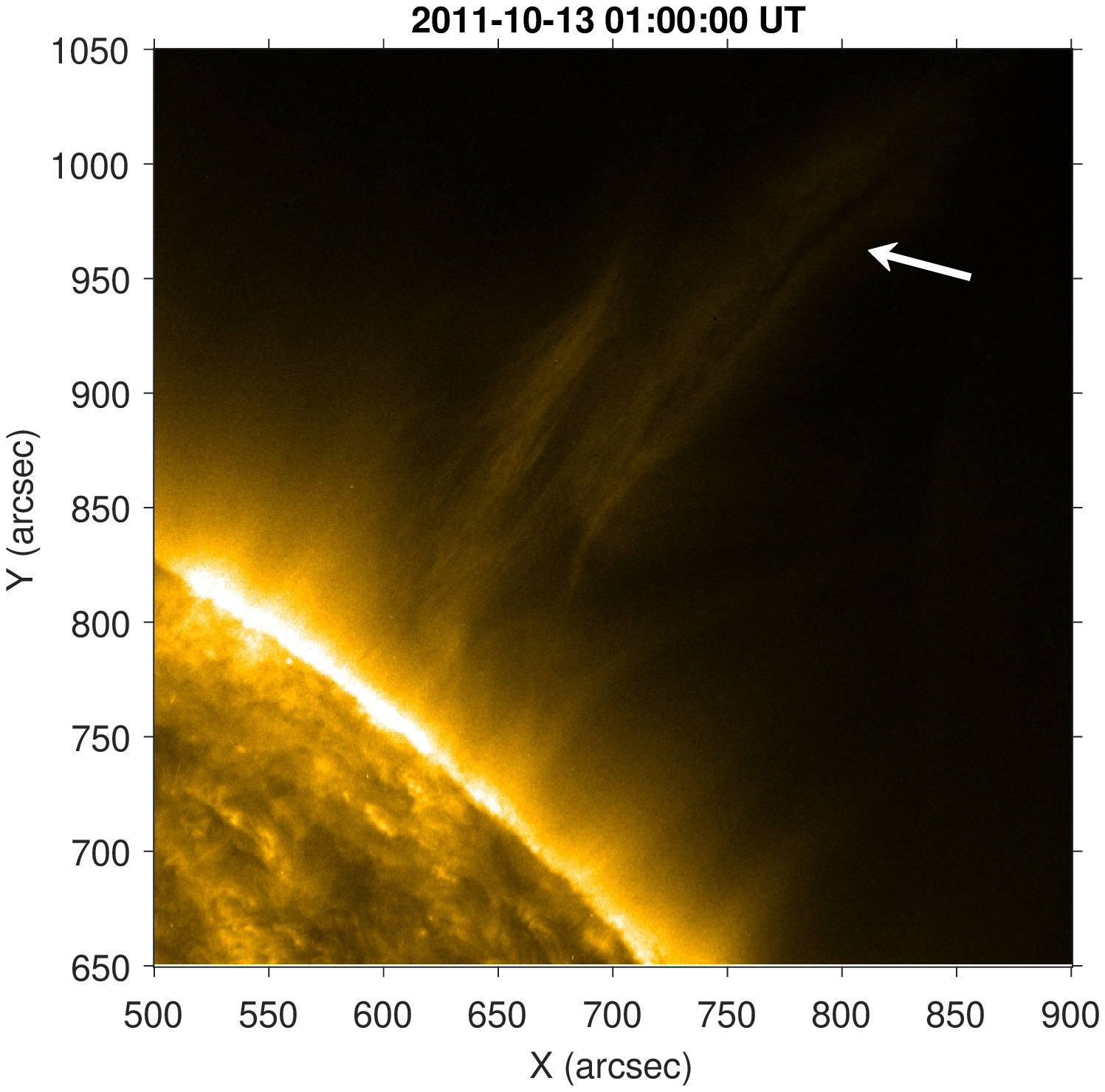}
\plotone{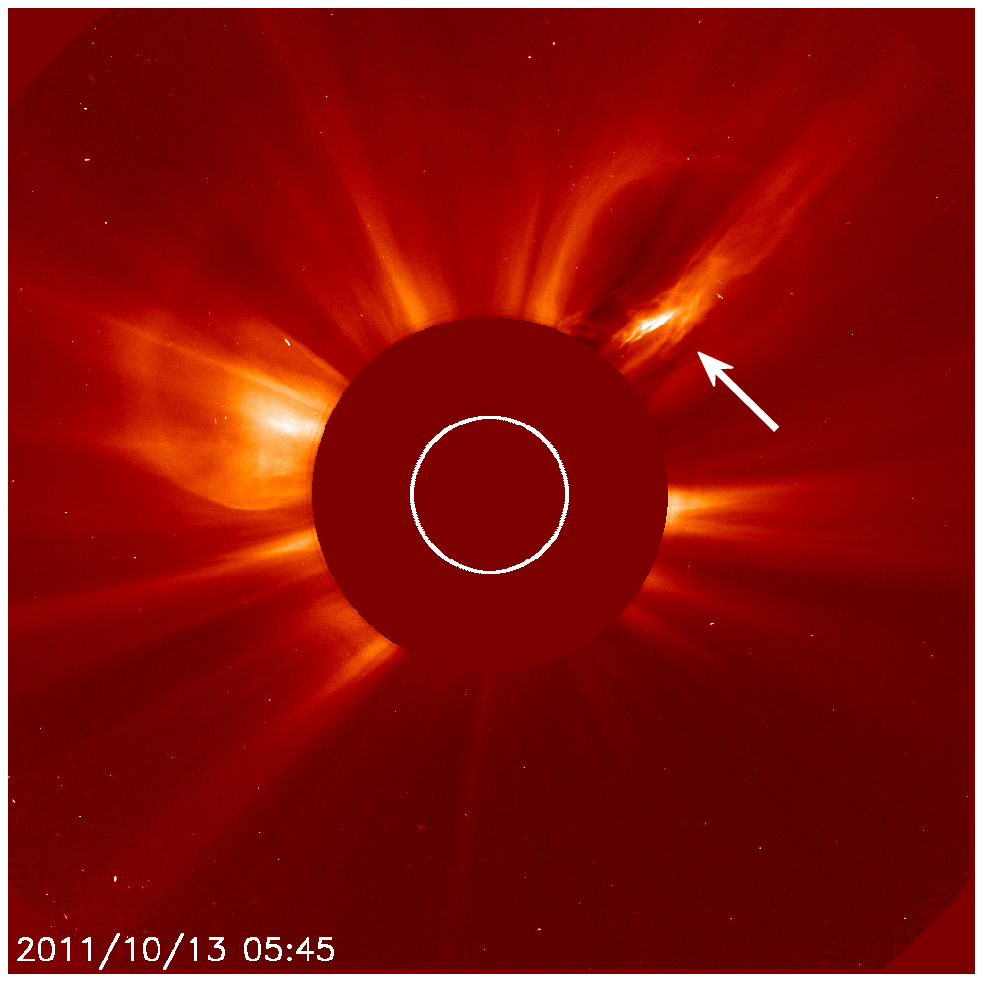}
\caption{Consecutive \emph{SDO}/AIA 171 {\AA} images (upper and lower left panels) showing evolution and eruption phases of an eruptive tornado during 2011 October 3-13. White arrows indicate the locations of the tornado axis at different times during its eruption. The lower right panel shows LASCO/C2 image of the corresponding CME at 05:45 UT, 2011 October 13. The white arrow again indicates a possible location of the upward ascending tornado.
The corresponding animation shows the \emph{SDO}/AIA 171 {\AA} images of the eruptive tornado from UT 2011 October 08T00:40 to 2011 October 13T05:50.}
(An animation of this figure is available.)\label{fig3}
\end{figure}

The lifetime of all detected prominence tornadoes ranges between 7 and 398 hr. Obviously, we could not trace all prominence tornadoes from the beginning to the end of their lifetimes. Some tornadoes disappeared into the far side of the Sun and we could not determine their lifetimes. Some tornadoes appeared from the far side and hence we had no information about their birth/start times. The average lifetime of all observed events is 126 hr, which is much longer than that (35 hr) obtained by \citet{Wedemeyer2013}. The number of tornadoes with lifetime $>$250 hr was 63 (17$\%$), the lifetime of 260 (72$\%$) tornadoes was under 150 hr and only 38 (11$\%$) tornadoes had a lifetime between 150 and 250 hr. Corresponding histogram is shown in Figure~\ref{fig1}. The figure shows a clear peak of tornado lifetime on 25-75 hr. Another small interesting peak is seen at 300-350 hr. It is unclear whether this peak has any real physical explanation, but almost all long lived tornadoes led to CMEs.

For those tornadoes that we traced from their formation to the end, the average lifetime is 88 hr. This is closer to the result of \citet{Wedemeyer2013}, but still more than twice longer. Panel (a) of Figure~\ref{fig2} shows the lifetime distribution of all traced tornadoes, while panels (b) and (c) show lifetimes of eruptive (which erupted as CMEs) and active tornadoes.

\section{Results} \label{sec:results}

After careful analysis of the observational data, we classified the prominence tornadoes into several categories. We excluded the tornadoes for which we were unable to study the full lifecycle (see also Section 2). Sometimes it was difficult to trace the tornadoes when they moved from the west limb toward the disk center. We could not trace 35 such tornadoes during the whole study.  Obviously, we were not able to trace the tornadoes that disappeared from the east limb into the far side of the Sun. There were 160 such events. Therefore, in total, 195 events from all detected tornadoes were excluded from further consideration. Consequently, the final phases of 166 prominence tornadoes were observed during the whole year. These 166 tornadoes can be classified into three categories: eruptive, active, and quasi-quiescent tornadoes (see Table 1). We will discuss each category separately.

\subsection{Eruptive tornadoes} \label{sec:results}

Tornadoes that become unstable and erupt at the final stage of their lifetimes are classified as eruptive. In turn, these eruptions lead to the destabilization of the hosting prominences, which finally eject as CMEs. We observed 80 eruptive tornadoes. The corresponding CMEs were identified in the LASCO CME catalog \citep{Yashiro2004}.

\begin{figure}
\epsscale{0.57}
\plotone{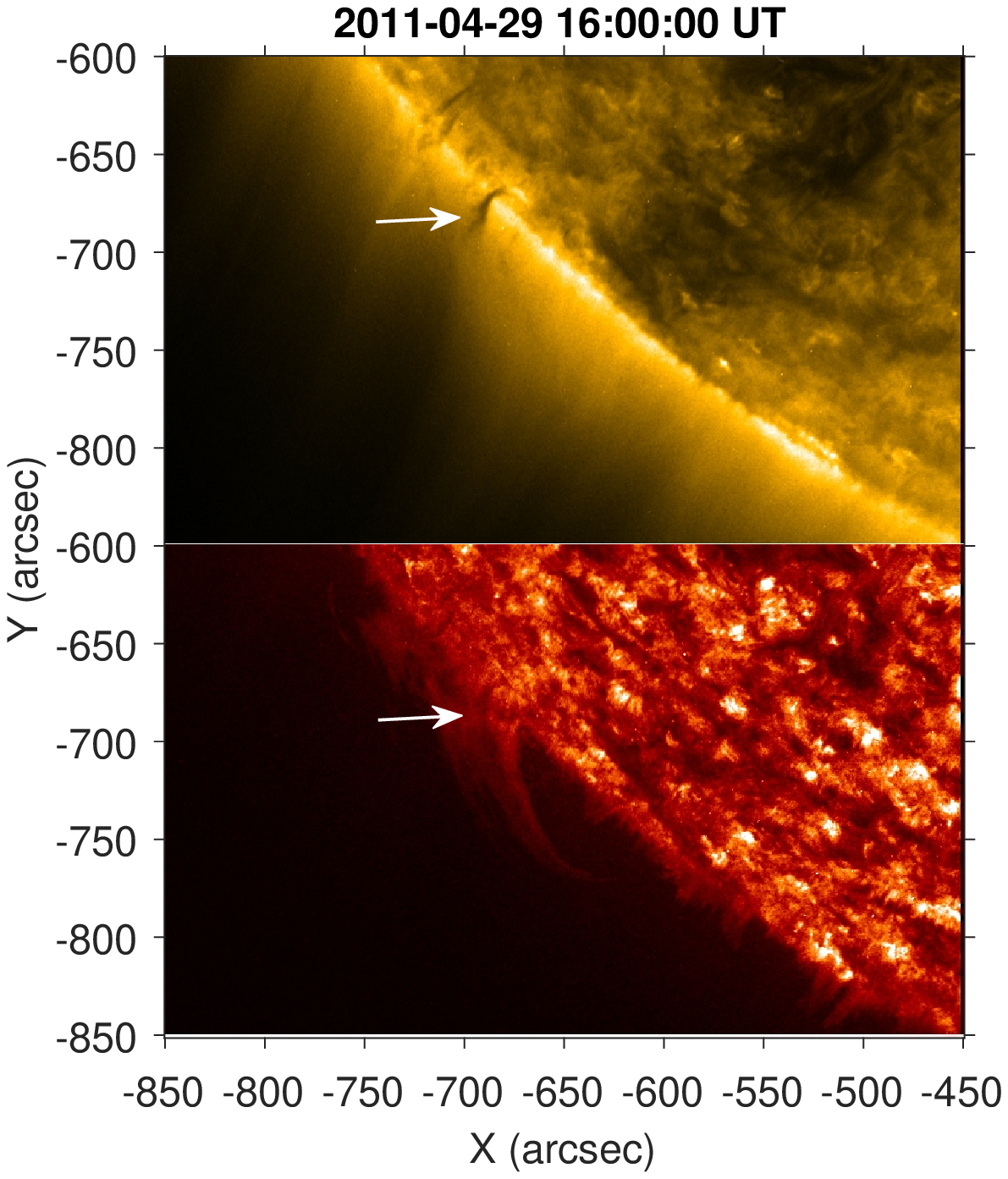}
\plotone{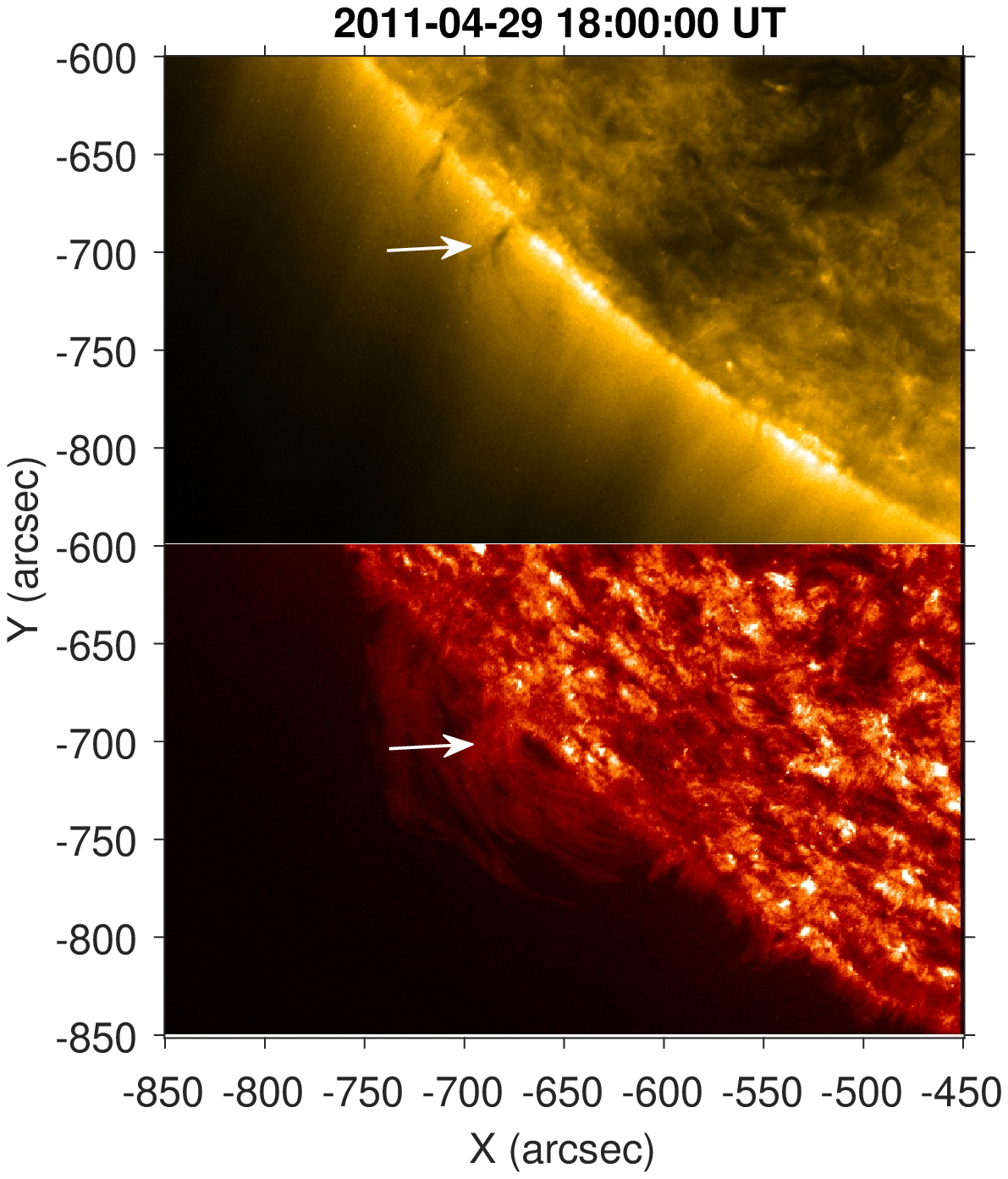}
\plotone{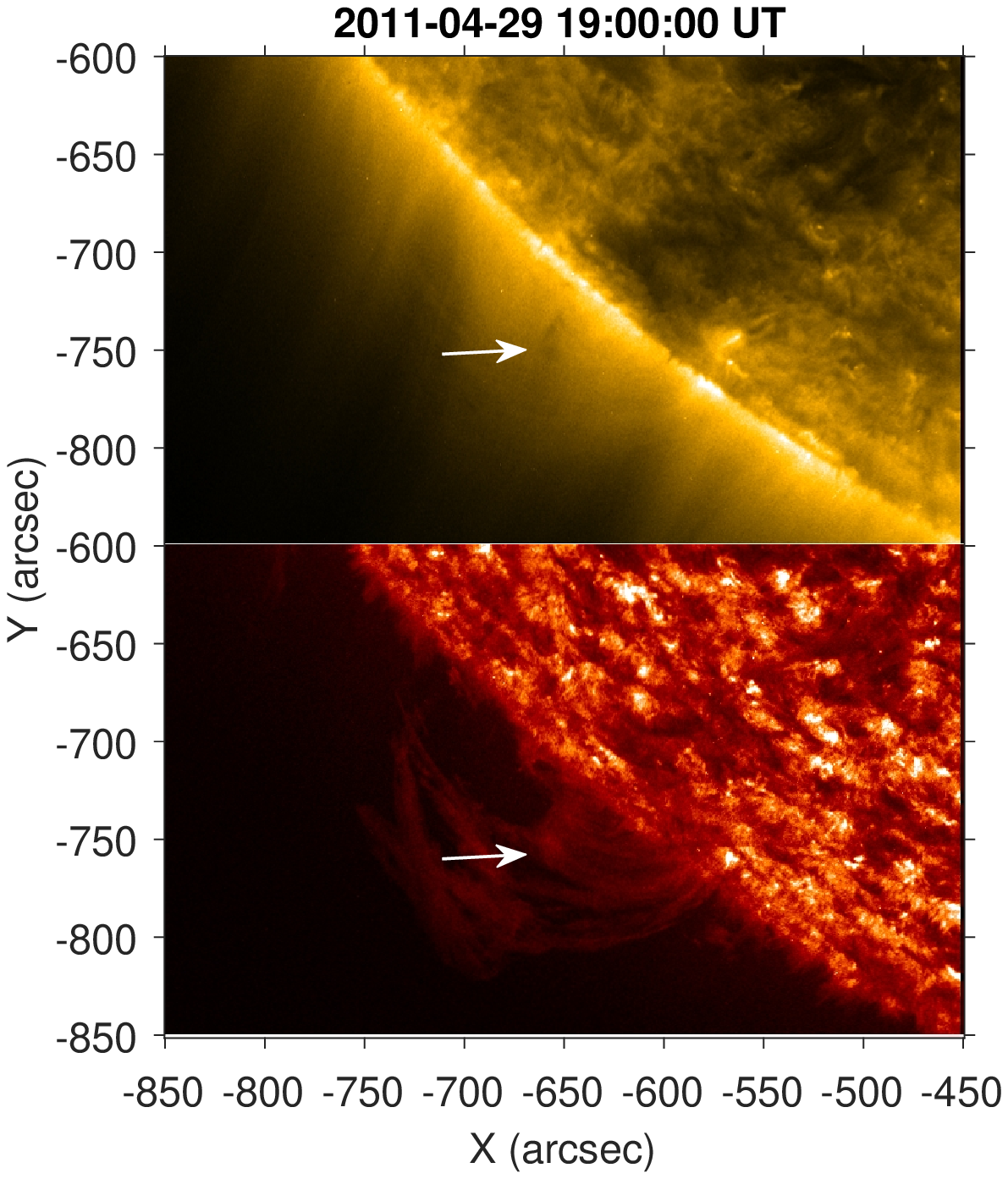}
\plotone{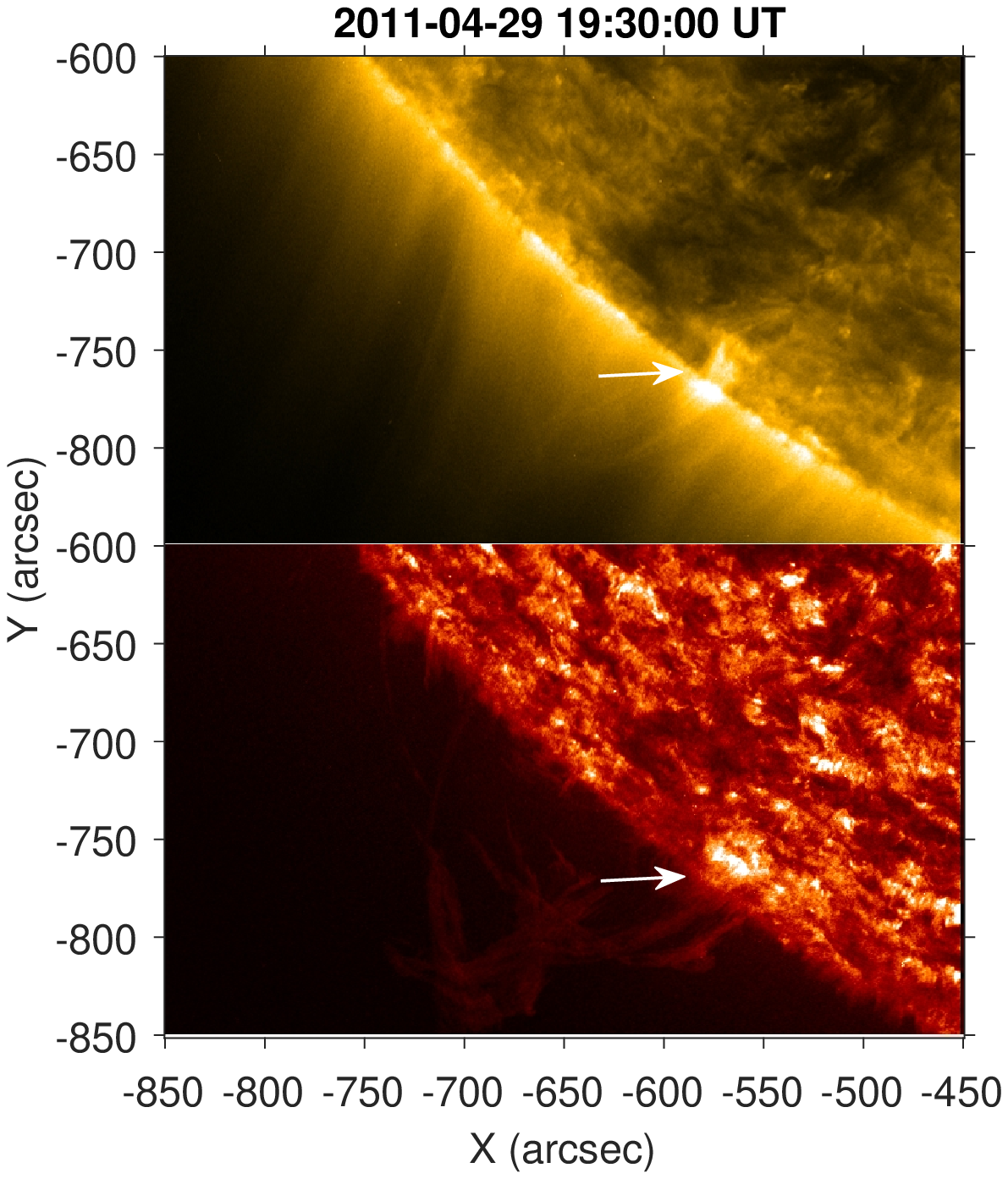}
\caption{Active tornado observed by \emph{SDO}/AIA 171 and 304 {\AA}. White arrows  in both line images show the location of the tornado.
The corresponding animation compares the \emph{SDO}/AIA 171 and 304 {\AA} images of the active tornado from UT 2011 April 27T15:06 to 2011 April 29T22:50.}
(An animation of this figure is available.)
\label{fig4}
\end{figure}

Figure~\ref{fig3} shows the evolution and final eruption phase of an eruptive tornado in the 171 {\AA} filter, which is a representative example for many eruptive tornadoes. The tornado was formed at 17:00 UT, October 3,  persisted for about 10 days and erupted at about 01:00 UT, October 13. The tornado rose up from the solar limb and finally erupted together with the corresponding prominence as a CME. Similar evolution and behavior are common characteristics for all eruptive tornadoes. The LASCO CME catalog reports the first appearance of the CME in the field of view of the C2 coronagraph at 02:03:24 UT, October 13 with a mean speed of 216 km s$^{-1}$. The CME appeared with an angular width of 125$^{\circ}$ in the northeast limb at a central position angle of 45$^{\circ}$ (see the lower right panel on Figure~\ref{fig3}).

\subsection{Active tornadoes} \label{sec:results}

The second category includes the tornadoes that produced some activity by the end of their lifetimes. Some of the tornadoes were associated with active filaments in event catalogs. Some of them led to failed CMEs when the eruption did not lead to a CME and drained back onto the solar surface. Some of the tornadoes were associated with quiescent prominences, but they have undergone some activity, such as transition flows which are connecting with hosting events. Tornadoes of this category often experienced some kind of influence from the surrounding medium.

\begin{figure}
\epsscale{0.58}
\plotone{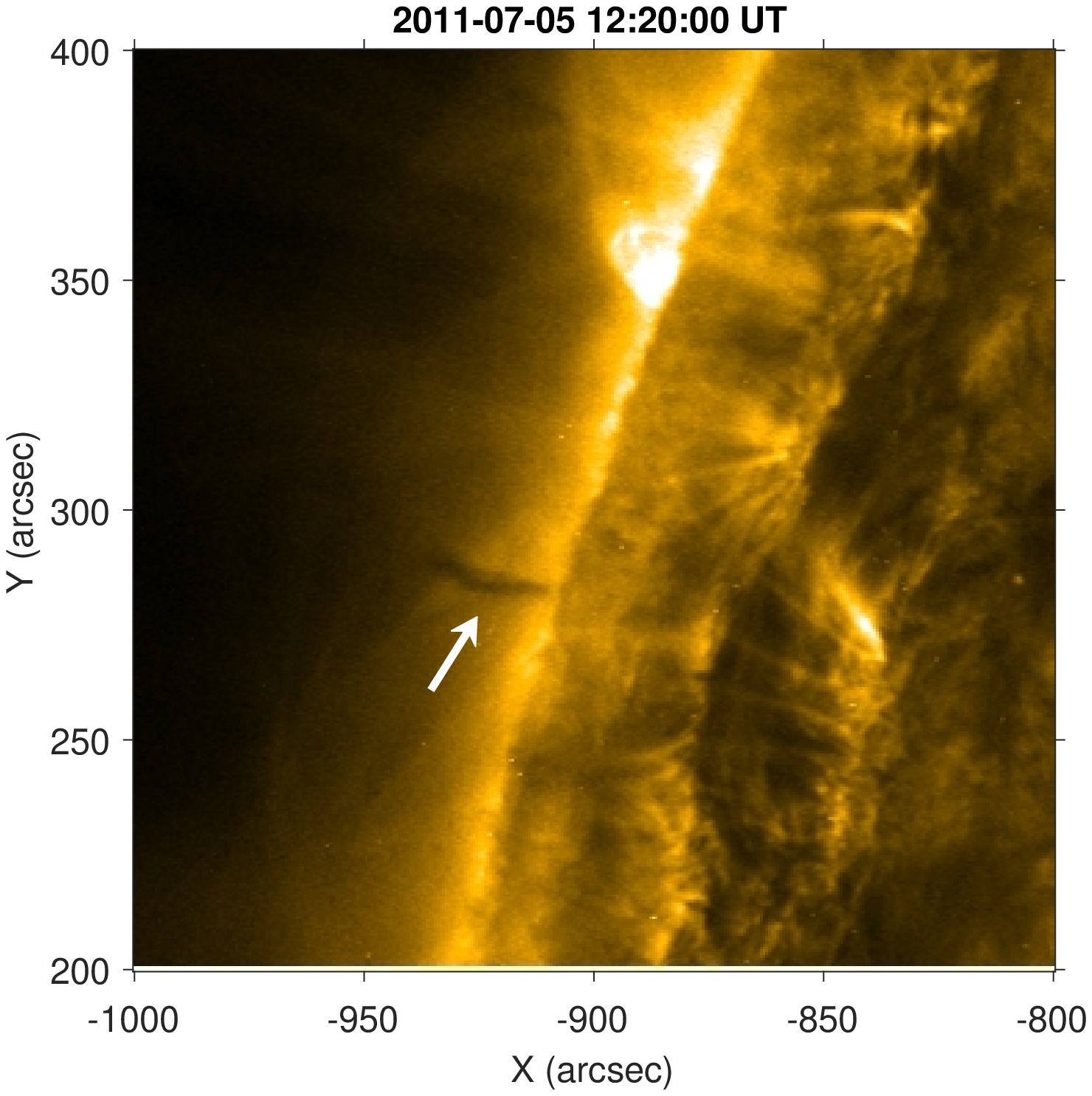}
\plotone{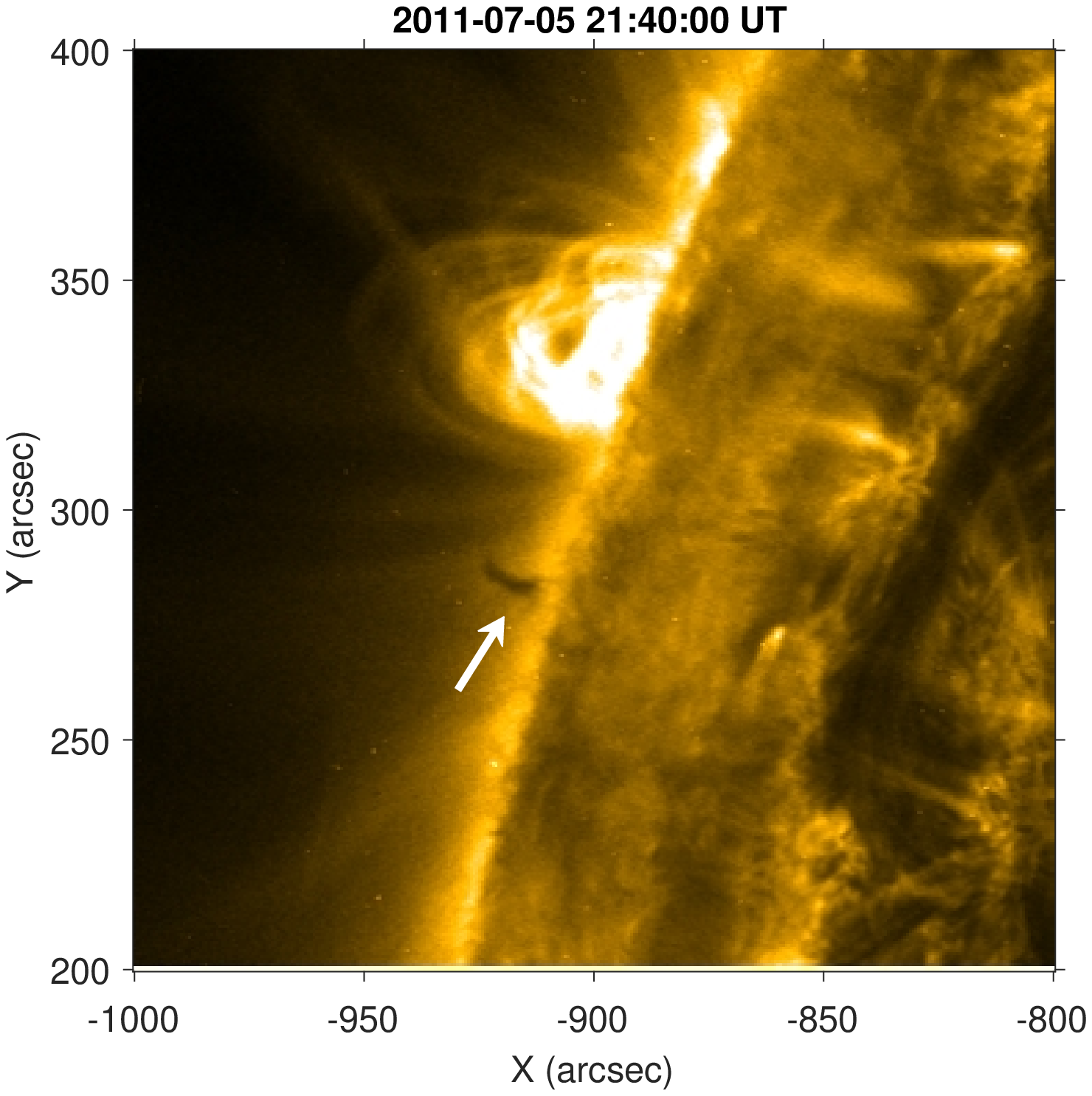}
\plotone{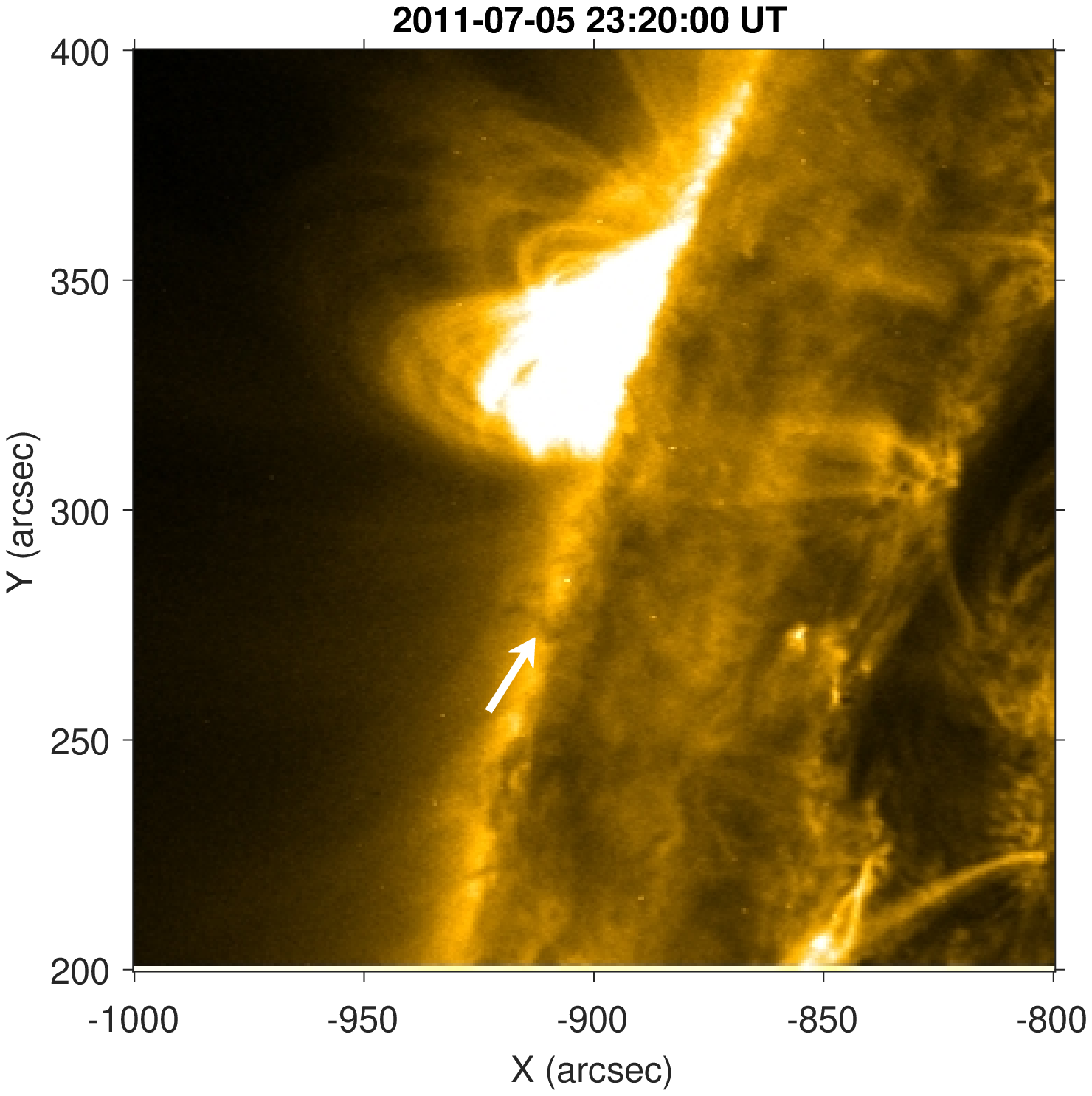}
\caption{Final evolution stage of a quasi-quiescent tornado in the AIA 171 {\AA} line. White arrows indicate the positions of the tornado.
The corresponding movie is available online. The animation shows the \emph{SDO}/AIA 171 {\AA} images of the active tornado from UT 2011-07-05T00:06 to 2011-07-6T02:58.}
(An animation of this figure is available.)
\label{fig5}
\end{figure}

An example of active tornadoes is shown in Figure~\ref{fig4}. In this case, we observed several active tornadoes in the 171 and 304 {\AA} filters. In the 171 {\AA} line, we see that the tornado group consisted of two clear neighboring structures (see, e.g., upper left panel corresponding to the time 16:00 UT, April 29). The tornadoes started to show some activity at about 18:00 UT, April 29, as they slowly rose up (see the upper right panel). The tornadoes consequently moved to the other side of the loops and disappeared near the surface. On the other hand, the 304 {\AA} line images clearly show how corresponding bright plasma was moving toward the other side of loops and caused brightening near the surface. Therefore, active tornadoes generally cause the destabilization of their hosting prominence and lead to  different types of activity such as a microflare and/or a failed CME. Out of 166 events, 83 were classified as active tornadoes.

\subsection{Quasi-quiescent tornadoes} \label{sec:results}

The third category includes the tornadoes, which are decreasing gradually in size and disappearing without any clear activity. There were only three events of this kind during the whole year. Figure~\ref{fig5} shows the evolution of the quasi-quiescent tornado during 05:00-23:00 UT on 2011 July 05. It is clearly seen in the figure how the tornado became smaller and disappeared in the 171 {\AA} line. The 304 {\AA} line also shows no clear activity in the hosting prominence. However, some clear brightening is seen in a nearby active region (northeast) after the tornado's disappearance. We can neither prove nor disprove a connection between the tornado and the active region brightening. Therefore, we call them quasi-quiescent tornadoes. Even though the number of these tornadoes is very small (only 3 out of 166), it raises some questions about the feasibility of the disappearance of the tornadoes without any activity. It is possible that we cannot detect the activity due to the spatial resolution or some other reason.

\section{Discussion and Conclusion} \label{sec:conslusion}

Prominence tornadoes, which are seen as dark structures in hot coronal lines, are usually associated with the prominence legs; therefore, their dynamics is closely connected with the evolution of the whole prominence itself. The rotational dynamics of the tornadoes is currently under debate. Some recent magnetic field measurements showed that the orientation of the field near prominence legs is primarily horizontal, i.e., parallel to the limb \citep{Schmieder2015,Levens2016a,Levens2016b,Levens2017}. The measurements could be subject to a 90$^{\circ}$ ambiguity; therefore, other configurations of magnetic field structure are also possible. However, the observations indicate the presence of significant twist inside the tornadoes as they appear to be vertically elongated structures. \citet{Gonzalez2016} detected a helical magnetic field structure near the legs of tornado-hosting prominence. Therefore, tornadoes most likely resemble twisted flux tubes or flux ropes. Twisted, non-potential magnetic configurations  suggest that they can efficiently store free magnetic energy, which is the main supply for the CMEs and eruptions \citep{Priest2002}. Furthermore, active regions with horizontal magnetic fields are likely to produce strong flares \citep{Kusano2012}. The magnetic field strength is generally around 15 G, but might reach 40-60 G in some places \citep{Levens2016b}. The twisted tubes usually undergo magnetic instabilities (e.g., kink instability) when the twist exceeds some threshold value \citep{Lundquist1951,Kuridze2013,Zaqarashvili2014}. Therefore, tornadoes may become unstable due to enhanced tension in the magnetic field. In this case, they may trigger an instability of their hosting prominence on larger spatial scales, which could lead to CMEs \citep{Wedemeyer2013}. On the other hand, tornadoes can also be unstable to the Kelvin-Helmholtz instability \citep{Zaqarashvili2010,Zaqarashvili2015}, which may lead to the heating of the structure and its consequent disappearance.

In order to study the stability properties of prominence tornadoes statistically, we used \emph{SDO}/AIA data from the year 2011. We carefully studied all tornadoes that appeared in the 171 {\AA} line during the whole year. In total, 361 events were detected, which include both individual tornadoes and tornado groups. However, we were not able to trace all of them until the end of their lives, as some tornadoes moved to the far side of the Sun due to the solar rotation and some of them moved toward the disk center where their identification was complicated. Finally, we selected 166 tornadoes which were tracked until the end of their lives confidently. Each of the tornadoes has been carefully observed, especially during the final phase of their evolution. We made a catalog of these 166 tornadoes, including their start and end times, starting solar coordinates and the lifetimes (see Table 2). Depending on their final fate, we classified the tornadoes into three categories. \textit{Eruptive tornadoes} lead to global instabilities of their hosting prominence and consequent CME's, which were found in the LASCO CME catalog \citep{Yashiro2004}. The appearance times of the CMEs are presented on the last column of Table 2. \textit{Active tornadoes} generally cause some apparent activity in their hosting prominences such as failed CMEs and/or plasma internal motion. These tornadoes sometimes also trigger microflares. \textit{Quasi-quiescent tornadoes} generally decay without showing any significant activity. The corresponding types of tornadoes are indicated in Table 2.

All observed tornadoes were associated with prominences. We have not investigated the prominences without tornadoes, but according to HEK there were 7529 prominences during the whole year of 2011. It is plausible that only one-third were seen near the limb, which leads to $\sim$ 2500 limb prominences per year.  \citet{Wang2010} also found that there are at least 10 limb prominences per day during the minimum phase of solar activity. Then using the total number of detected tornadoes during 2011 (361) one can conclude that roughly 10-15\% of prominences host tornadoes.

From the selected 166 tornadoes, 80 ($48 \%$) were eruptive, i.e., they caused CMEs, 83 ($50 \%$) were active, and only 3 ($2 \%$) were quasi-quiescent, i.e., they disappeared without any noticeable activity. This means that almost all tornado-hosting prominences show some activity in the form of eruption, failed eruption, or internal instability. On the other hand, almost half of the tornado-hosting prominences erupted as CMEs. This is a very interesting statistical result, which could play a significant role in space weather predictions. If a tornado-like structure appears near a prominence leg, then it is likely to erupt as a CME with $50 \%$ probability.
The mean lifetime of eruptive tornadoes is between 25 and 75 hr, which means that the CME initiation may be expected to occur 1-3 days after the formation of a tornado-like structure in a prominence. One can predict the approximate location of the prominence on the solar disk, which will help to predict the expected angle of CME eruption. This is one of the key issues in space weather. Future observations will show more details of the connection between tornado-hosting prominences and CMEs.

CMEs that lead to shocks produce type II radio bursts, while flare-generated electron beams cause type III radio bursts \citep{White2007}. It would be interesting to study whether prominence tornadoes have any radio emission signatures associated with them. The considered year (2011) was close to the peak of solar activity. It would be interesting to study how the results of this work change at solar activity minimum, i.e., what is the effect of the solar cycle on our statistical results. We plan to perform 3D MHD simulations of tornado and prominence destabilization, particulary focusing on the effect of horizontal magnetic field as in \citet{Kusano2012}.

{\bf Acknowledgements} The work was supported by Georgian Shota Rustaveli National Science Foundation project DI-2016-17 and by the Austrian Science Fund  (FWF, project 30695-N27). B.M.S. acknowledges the support by the FWF projects P25640-N27, S11606-N16, also Leverhulme Trust grant IN-2014-016. D.K. has received funding from the S\^er Cymru II Part-funded by the European Regional Development Fund through the Welsh Government.

\clearpage

\appendix

A catalog of prominence tornadoes during the year 2011 is presented here. First and second columns stand for start and end times of tornadoes. Third and fourth columns show initial X and Y  coordinates of the tornadoes. Fifth column indicates tornado type. Sixth and seventh columns denote life times of tornadoes and corresponding CME appearance times, respectively.

\startlongtable
\begin{deluxetable*}{llccccc}
\tablecaption{A catalog of prominence tornadoes during the year 2011.\label{tab2}}
\tablewidth{700pt}
\tabletypesize{\scriptsize}
\tablehead{
\colhead{Start Time} & \colhead{End Time} &
\colhead{Start X} & \colhead{Start Y} &
\colhead{Type} & \colhead{Lifetime (Hr)} &
\colhead{CME appearance Time}
}
\startdata
31.12.2010 05:00 & 02.01.2011 15:00 & 524 & -829 & CME & 58 & 1/2/2011 15:24:05 \\
27.12.2010 13:00 & 01.01.2011 14:30 & -425 & 775 & Active & 126 & \nodata \\
30.12.2010 08:00 & 01.01.2011 14:00 & -332 & 870 & Active & 54 & \nodata \\
01.01.2011 13:00 & 02.01.2011 05:00 & 528 & -674 & CME & 16 & 1/2/2011 15:24:05 \\
29.12.2010 18:00 & 02.01.2011 23:00 & 548 & 806 & Active & 101 & \nodata \\
17.01.2011 14:00 & 18.01.2011 03:00 & 362 & 861 & Active & 13 & \nodata \\
17.01.2011 13:00 & 22.01.2011 21:00 & 300 & 560 & CME & 128 & 1/23/2011 0:48:05 \\
19.01.2011 04:00 & 20.01.2011 15:00 & -897 & 354 & Active & 35 & \nodata \\
17.01.2011 20:00 & 20.01.2011 16:00 & 465 & 836 & CME & 68 & 1/20/2011 18:48:06 \\
20.01.2011 20:00 & 22.01.2011 12:00 & 632 & 729 & CME & 40 & 1/22/2011 13:36:06 \\
30.01.2011 05:00 & 03.02.2011 17:00 & -865 & 422 & Active & 108 & \nodata \\
05.02.2011 16:00 & 08.02.2011 19:00 & -791 & 532 & CME & 75 & 2/9/2011 0:12:06 \\
07.02.2011 13:00 & 08.02.2011 19:00 & -503 & 822 & CME & 30 & 2/9/2011 0:12:06 \\
06.02.2011 21:00 & 07.02.2011 04:00 & -664 & 705 & Active & 7 & \nodata \\
25.01.2011 07:00 & 28.01.2011 03:00 & -870 & -416 & CME & 68 & 1/28/2011 5:00:07 \\
05.02.2011 10:00 & 10.02.2011 21:00 & -762 & -584 & CME & 131 & 2/11/2011 1:36:05 \\
09.02.2011 21:00 & 10.02.2011 11:00 & -775 & -568 & Active & 14 & \nodata \\
11.02.2011 07:00 & 11.02.2011 22:00 & -765 & -583 & Active & 15 & \nodata \\
29.01.2011 13:00 & 31.01.2011 03:00 & 548 & -786 & Active & 38 & \nodata \\
24.01.2011 13:00 & 25.01.2011 03:00 & 723 & -515 & Active & 14 & \nodata \\
21.02.2011 09:00 & 25.02.2011 06:00 & -767 & 569 & CME & 93 & 2/25/2011 6:12:06 \\
24.02.2011 05:00 & 25.02.2011 06:00 & -505 & 793 & CME & 25 & 2/25/2011 6:12:06 \\
27.02.2011 05:00 & 28.02.2011 08:00 & -548 & 804 & Active & 27 & \nodata \\
17.02.2011 17:00 & 25.02.2011 19:00 & -438 & 858 & CME & 194 & 2/25/2011 20:24:06 \\
11.02.2011 01:00 & 17.02.2011 11:00 & -530 & -800 & Active & 154 & \nodata \\
20.02.2011 07:00 & 01.03.2011 23:00 & -925 & -206 & CME & 232 & 3/2/2011 2:24:07 \\
20.02.2011 07:00 & 01.03.2011 23:00 & -874 & -382 & CME & 232 & 3/2/2011 2:24:07 \\
04.03.2011 07:00 & 05.03.2011 19:00 & -885 & 370 & CME & 36 & 3/5/2011 20:54:05 \\
08.03.2011 21:00 & 20.03.2011 03:00 & -882 & 328 & Active & 270 & \nodata \\
11.03.2011 19:00 & 13.03.2011 19:00 & -693 & 649 & Active & 48 & \nodata \\
10.03.2011 13:00 & 12.03.2011 17:00 & -947 & -61 & CME & 52 & 3/12/2011 17:12:05 \\
12.03.2011 08:00 & 14.03.2011 02:00 & -697 & 663 & Active & 42 & \nodata \\
15.03.2011 17:00 & 18.03.2011 03:00 & -95 & 902 & Active & 58 & \nodata \\
01.03.2011 22:00 & 03.03.2011 02:00 & 724 & 589 & Active & 28 & \nodata \\
05.03.2011 11:00 & 06.03.2011 13:00 & -669 & -681 & CME & 26 & 3/6/2011 5:12:06 \\
09.03.2011 07:00 & 10.03.2011 03:00 & -840 & -441 & Active & 20 & \nodata \\
14.03.2011 03:00 & 19.03.2011 21:00 & -544 & -781 & CME & 138 & 3/19/2011 21:24:11 \\
01.03.2011 21:00 & 05.03.2011 05:00 & 788 & -513 & Active & 80 & \nodata \\
15.03.2011 01:00 & 17.03.2011 23:00 & -858 & 412 & Active & 70 & \nodata \\
19.03.2011 19:00 & 21.03.2011 23:00 & -631 & 697 & Active & 52 & \nodata \\
19.03.2011 17:00 & 27.03.2011 21:00 & -433 & 840 & Active & 196 & \nodata \\
25.03.2011 11:00 & 29.03.2011 19:00 & -559 & 760 & CME & 104 & 3/29/2011 20:36:07 \\
22.03.2011 13:00 & 24.03.2011 15:00 & -481 & 812 & Active & 50 & \nodata \\
23.03.2011 03:00 & 24.03.2011 03:00 & -670 & -676 & CME & 24 & 3/24/2011 4:36:07 \\
27.03.2011 15:00 & 09.04.2011 11:00 & -665 & -674 & Active & 308 & \nodata \\
11.03.2011 17:00 & 19.03.2011 11:00 & 57 & -343 & CME & 186 & 3/19/2011 12:12:06 \\
30.03.2011 17:00 & 06.04.2011 14:00 & -390 & 882 & Active & 141 & \nodata \\
04.04.2011 12:00 & 05.04.2011 21:00 & -949 & 160 & Active & 33 & \nodata \\
08.04.2011 05:00 & 09.04.2011 00:00 & -960 & -48 & Active & 19 & \nodata \\
05.04.2011 07:00 & 06.04.2011 23:00 & -637 & 723 & Active & 40 & \nodata \\
02.04.2011 02:00 & 02.04.2011 15:00 & 614 & 742 & Active & 13 & \nodata \\
04.04.2011 05:00 & 12.04.2011 11:00 & -72 & 790 & Active & 198 & \nodata \\
10.04.2011 21:00 & 15.04.2011 03:00 & 368 & 844 & CME & 102 & 4/15/2011 13:36:05 \\
30.03.2011 19:00 & 11.04.2011 23:00 & -929 & 234 & Active & 292 & \nodata \\
04.04.2011 03:00 & 17.04.2011 15:00 & -734 & -590 & CME & 324 & 4/17/2011 19:36:05 \\
10.04.2011 03:00 & 14.04.2011 23:00 & -752 & -583 & Active & 116 & \nodata \\
11.04.2011 13:00 & 13.04.2011 07:00 & 696 & -669 & CME & 42 & 4/13/2011 16:48:05 \\
17.04.2011 15:00 & 20.04.2011 21:00 & -941 & -70 & Active & 78 & \nodata \\
19.04.2011 05:00 & 19.04.2011 22:00 & -808 & 471 & Active & 17 & \nodata \\
17.04.2011 05:00 & 20.04.2011 15:00 & -701 & 659 & CME & 82 & 4/20/2011 5:12:07 \\
15.04.2011 10:00 & 17.04.2011 07:00 & -501 & 821 & CME & 45 & 4/17/2011 11:48:06 \\
19.04.2011 11:00 & 21.04.2011 07:00 & -429 & 841 & Active & 44 & \nodata \\
27.04.2011 07:00 & 29.04.2011 07:00 & 667 & 636 & Active & 48 & \nodata \\
24.04.2011 23:00 & 26.04.2011 23:00 & -659 & -670 & CME & 48 & 4/27/2011 7:48:06 \\
27.04.2011 15:00 & 29.04.2011 19:00 & -679 & -637 & Active & 52 & \nodata \\
21.04.2011 11:00 & 25.04.2011 09:00 & -459 & -818 & Active & 94 & \nodata \\
29.04.2011 23:00 & 02.05.2011 07:00 & 662 & -687 & CME & 56 & 5/2/2011 7:36:07 \\
11.05.2011 19:00 & 12.05.2011 19:00 & -906 & 256 & Active & 24 & \nodata \\
13.05.2011 03:00 & 14.05.2011 11:00 & -765 & 535 & Active & 32 & \nodata \\
03.05.2011 19:00 & 05.05.2011 11:00 & -643 & 693 & Active & 40 & \nodata \\
09.05.2011 01:00 & 09.05.2011 23:00 & -645 & 669 & Active & 22 & \nodata \\
27.04.2011 15:00 & 02.05.2011 17:00 & 202 & 779 & Active & 122 & \nodata \\
03.05.2011 17:00 & 04.05.2011 15:00 & -895 & -315 & Active & 22 & \nodata \\
08.05.2011 05:00 & 13.05.2011 17:00 & -19 & -467 & Active & 132 & \nodata \\
19.05.2011 21:00 & 25.05.2011 11:00 & -781 & 511 & Active & 134 & \nodata \\
27.05.2011 01:00 & 10.06.2011 07:00 & -718 & 599 & CME & 342 & 6/10/2011 13:25:46 \\
15.05.2011 19:00 & 18.05.2011 19:00 & -456 & 824 & CME & 72 & 5/18/2011 21:24:09 \\
19.05.2011 07:00 & 19.05.2011 23:00 & -595 & 723 & CME & 16 & 5/20/2011 3:12:09 \\
30.05.2011 07:00 & 31.05.2011 11:00 & -564 & 753 & CME & 28 & 5/31/2011 14:12:06 \\
18.05.2011 17:00 & 21.05.2011 15:00 & 742 & 552 & Active & 70 & \nodata \\
16.05.2011 05:00 & 18.05.2011 04:00 & 786 & 434 & CME & 47 & 5/18/2011 6:24:06 \\
23.05.2011 17:00 & 25.05.2011 21:00 & -932 & -68 & Active & 52 & \nodata \\
25.05.2011 01:00 & 27.05.2011 23:00 & -898 & -266 & Active & 70 & \nodata \\
30.05.2011 17:00 & 12.06.2011 15:00 & -931 & 79 & CME & 310 & 6/12/2011 14:48:06 \\
29.05.2011 11:00 & 01.06.2011 07:00 & -755 & -547 & Active & 68 & \nodata \\
25.05.2011 11:00 & 26.05.2011 07:00 & -487 & -803 & CME & 20 & 5/26/2011 4:36:05 \\
15.05.2011 19:00 & 17.05.2011 21:00 & 843 & -267 & Active & 50 & \nodata \\
09.06.2011 06:00 & 14.06.2011 05:00 & -960 & 78 & CME & 119 & 6/14/2011 6:12:05 \\
31.05.2011 03:00 & 05.06.2011 05:00 & -802 & 476 & CME & 122 & 6/5/2011 3:44:23 \\
13.06.2011 07:00 & 21.06.2011 03:00 & -794 & 494 & CME & 188 & 6/21/2011 3:16:10 \\
14.06.2011 11:00 & 23.06.2011 23:00 & -629 & 694 & Active & 228 & \nodata \\
12.06.2011 03:00 & 12.06.2011 17:00 & 566 & 753 & Active & 15 & \nodata \\
14.06.2011 09:00 & 14.06.2011 19:00 & 809 & 489 & CME & 10 & 6/14/2011 18:36:05 \\
05.06.2011 09:00 & 06.06.2011 07:00 & -876 & -316 & Active & 22 & \nodata \\
03.06.2011 21:00 & 05.06.2011 11:00 & -800 & -484 & CME & 38 & 6/5/2011 16:59:57 \\
02.06.2011 03:00 & 05.06.2011 11:00 & -668 & -653 & CME & 80 & 6/5/2011 16:59:57 \\
10.06.2011 15:00 & 12.06.2011 07:00 & 815 & -263 & Active & 40 & \nodata \\
13.06.2011 15:00 & 15.06.2011 11:00 & 698 & -335 & CME & 44 & 6/15/2011 13:36:21 \\
21.05.2011 05:00 & 06.06.2011 03:00 & -759 & -544 & CME & 382 & 6/6/2011 7:30:04 \\
21.05.2011 19:00 & 06.06.2011 09:00 & -637 & -685 & CME & 374 & 6/6/2011 7:30:04 \\
12.06.2011 11:00 & 15.06.2011 17:00 & 500 & -675 & CME & 78 & 6/15/2011 13:36:21 \\
19.06.2011 05:00 & 23.06.2011 12:00 & -900 & 264 & Active & 103 & \nodata \\
19.06.2011 12:00 & 23.06.2011 12:00 & -814 & 516 & Active & 96 & \nodata \\
25.06.2011 03:00 & 28.06.2011 21:00 & -646 & 674 & Active & 90 & \nodata \\
15.06.2011 22:00 & 17.06.2011 11:00 & 646 & 658 & CME & 37 & 6/17/2011 11:00:05 \\
16.06.2011 19:00 & 17.06.2011 17:00 & -771 & -530 & Active & 22 & \nodata \\
19.06.2011 19:00 & 21.06.2011 09:00 & -653 & -665 & CME & 38 & 6/21/2011 8:48:05 \\
05.07.2011 05:00 & 05.07.2011 23:00 & -900 & 290 & quasi-quiescent & 18 & \nodata \\
08.07.2011 09:00 & 10.07.2011 07:00 & -675 & 665 & CME & 46 & 7/10/2011 12:00:05 \\
13.07.2011 17:00 & 17.07.2011 05:00 & -667 & 670 & CME & 84 & 7/17/2011 6:12:05 \\
11.07.2011 05:00 & 13.07.2011 21:00 & -581 & -735 & Active & 64 & \nodata \\
07.07.2011 05:00 & 09.07.2011 11:00 & 505 & -742 & Active & 54 & \nodata \\
17.07.2011 17:00 & 19.07.2011 22:00 & -436 & 843 & Active & 47 & \nodata \\
31.07.2011 22:00 & 01.08.2011 17:00 & -923 & 220 & Active & 19 & \nodata \\
07.08.2011 05:00 & 11.08.2011 15:00 & -660 & 681 & Active & 106 & \nodata \\
12.08.2011 19:00 & 17.08.2011 03:00 & -415 & 852 & CME & 104 & 8/17/2011 4:00:06 \\
14.08.2011 19:00 & 16.08.2011 07:00 & -715 & 624 & CME & 36 & 8/16/2011 10:12:06 \\
10.08.2011 17:00 & 14.08.2011 14:00 & 140 & -787 & Active & 93 & \nodata \\
22.08.2011 11:00 & 31.08.2011 09:00 & -681 & 664 & Active & 214 & \nodata \\
17.08.2011 18:00 & 19.08.2011 03:00 & -587 & 752 & quasi-quiescent & 9 & \nodata \\
18.08.2011 11:00 & 20.08.2011 23:00 & -343 & 885 & CME & 60 & 8/21/2011 1:36:20 \\
25.08.2011 07:00 & 30.08.2011 11:00 & 471 & 456 & Active & 124 & \nodata \\
30.08.2011 15:00 & 02.09.2011 07:00 & -782 & -542 & CME & 64 & 9/2/2011 0:36:06 \\
12.09.2011 23:00 & 14.09.2011 11:00 & -954 & -30 & Active & 36 & \nodata \\
31.08.2011 13:00 & 10.09.2011 03:00 & -840 & 420 & CME & 206 & 9/10/2011 3:12:09 \\
02.09.2011 23:00 & 04.09.2011 03:00 & -840 & 466 & Active & 28 & \nodata \\
13.09.2011 13:00 & 14.09.2011 13:00 & -676 & 679 & CME & 24 & 9/14/2011 12:24:06 \\
10.09.2011 03:00 & 13.09.2011 13:00 & -480 & 828 & CME & 82 & 9/13/2011 16:12:05 \\
14.09.2011 03:00 & 15.09.2011 15:00 & -546 & -788 & Active & 36 & \nodata \\
02.09.2011 05:00 & 06.09.2011 07:00 & 500 & -748 & CME & 98 & 9/6/2011 5:36:06 \\
10.09.2011 11:00 & 12.09.2011 13:00 & 606 & -737 & Active & 50 & \nodata \\
17.09.2011 01:00 & 19.09.2011 06:00 & -950 & 110 & Active & 53 & \nodata \\
03.10.2011 17:00 & 13.10.2011 01:00 & 157 & 667 & CME & 224 & 9/13/2011 1:25:52 \\
03.10.2011 05:00 & 04.10.2011 13:00 & -810 & -519 & Active & 32 & \nodata \\
08.10.2011 00:00 & 08.10.2011 18:00 & -802 & -522 & Active & 18 & \nodata \\
04.10.2011 05:00 & 05.10.2011 11:00 & -645 & -708 & CME & 30 & 10/5/2011 9:12:09 \\
05.10.2011 05:00 & 06.10.2011 03:00 & 633 & -695 & CME & 22 & 10/6/2011 5:24:06 \\
18.10.2011 03:00 & 22.10.2011 17:00 & 4 & 812 & CME & 110 & 10/22/2011 17:48:05 \\
24.10.2011 07:00 & 25.10.2011 05:00 & -958 & -122 & CME & 22 & 10/25/2011 6:24:05 \\
21.10.2011 11:00 & 24.10.2011 21:00 & -881 & -403 & CME & 82 & 10/24/2011 20:36:07 \\
03.11.2011 17:00 & 05.11.2011 03:00 & -968 & 45 & Active & 34 & \nodata \\
15.11.2011 15:00 & 26.11.2011 09:00 & -970 & 14 & Active & 258 & \nodata \\
08.11.2011 09:00 & 22.11.2011 19:00 & -885 & 401 & CME & 346 & 11/22/2011 20:57:31 \\
11.11.2011 13:00 & 14.11.2011 11:00 & -651 & 717 & CME & 70 & 11/14/2011 15:48:06 \\
29.10.2011 05:00 & 04.11.2011 21:00 & -13 & 718 & CME & 160 & 11/5/2011 0:36:05 \\
04.11.2011 07:00 & 05.11.2011 03:00 & 769 & -567 & CME & 20 & 11/5/2011 4:00:05 \\
26.11.2011 01:00 & 28.11.2011 07:00 & -822 & 525 & CME & 54 & 11/28/2011 6:36:05 \\
28.11.2011 13:00 & 29.11.2011 21:00 & 752 & 616 & Active & 32 & \nodata \\
24.11.2011 09:00 & 27.11.2011 17:00 & 498 & 251 & CME & 80 & 11/27/2011 20:00:07 \\
17.11.2011 19:00 & 20.11.2011 15:00 & 733 & 216 & CME & 68 & 11/20/2011 14:12:05 \\
05.12.2011 07:00 & 05.12.2011 19:00 & -948 & 222 & Active & 12 & \nodata \\
05.12.2011 23:00 & 11.12.2011 07:00 & -894 & 384 & CME & 128 & 12/11/2011 10:12:06 \\
29.11.2011 23:00 & 04.12.2011 03:00 & -796 & 543 & CME & 100 & 12/4/2011 6:00:05 \\
14.12.2011 05:00 & 18.12.2011 07:00 & -730 & 644 & CME & 98 & 12/18/2011 17:48:05 \\
04.12.2011 03:00 & 09.12.2011 23:00 & -578 & 777 & CME & 140 & 12/10/2011 7:24:06 \\
15.12.2011 01:00 & 18.12.2011 11:00 & -438 & 870 & quasi-quiescent & 82 & \nodata \\
29.11.2011 09:00 & 02.12.2011 11:00 & 355 & 883 & CME & 74 & 12/2/2011 17:24:05 \\
02.12.2011 05:00 & 03.12.2011 15:00 & -819 & -526 & CME & 34 & 12/3/2011 16:36:06 \\
15.12.2011 01:00 & 16.12.2011 09:00 & -812 & -544 & Active & 32 & \nodata \\
10.12.2011 13:00 & 24.12.2011 11:00 & -506 & -831 & Active & 334 & \nodata \\
22.12.2011 11:00 & 23.12.2011 23:00 & -821 & 519 & CME & 36 & 12/23/2011 20:48:07 \\
26.12.2011 11:00 & 28.12.2011 13:00 & 378 & 611 & CME & 50 & 12/28/2011 16:12:06 \\
31.12.2011 11:00 & 15.01.2012 03:00 & -914 & -344 & Active & 352 & \nodata \\
17.12.2011 01:00 & 18.12.2011 07:00 & -837 & -499 & CME & 30 & 12/19/2011 1:25:52 \\
31.12.2011 15:00 & 01.01.2012 19:00 & -650 & -730 & Active & 28 & \nodata \\
15.12.2011 15:00 & 16.12.2011 17:00 & 839 & -477 & CME & 26 & 12/16/2011 20:48:05 \\
\enddata
\end{deluxetable*}

\end{document}